\documentstyle[a4p,11pt,epsfig]{article}
\def\Journal#1#2#3#4{{#1}{\bf#2} (#4) #3}

\def\NIMA{{\sl Nucl.\ Inst.\ Meth.} \bf A}
\def\NPB{{\sl Nucl.\ Phys.} \bf B}
\def\PLB{{\sl Phys.\ Lett.} \bf B}
\def\PRL{\sl Phys.\ Rev.\ Lett.\ }
\def\PRD{{\sl Phys.\ Rev.} \bf D}
\def\ZPC{{\sl Z.\ Phys.} \bf C}
\def\EPJ{{\sl Eur.\ Phys.\ J.} \bf C}
\newcommand{\etal} {{\it et al.,} }
\newcommand{\plb}{\mbox{$\langle P_L^{\Lambda_{\mathrm{b}}} \rangle$}}
\newcommand{\pl}{\mbox{$\langle P_L \rangle$}}
\newcommand{\plbn}{\mbox{$P_L^{\Lambda_{\mathrm{b}}}$}}
\newcommand{\lb}{\mbox{$\Lambda_{\mathrm{b}}$}}
\newcommand{\lc}{\mbox{$\Lambda_{\mathrm{c}}$}}
\newcommand{\lcp}{\mbox{$\Lambda_{\mathrm{c}}^+$}}
\newcommand{\xb}{\mbox{$\Xi_{\mathrm{b}}$}}
\newcommand{\xc}{\mbox{$\Xi_{\mathrm{c}}$}}
\newcommand{\lbz}{\mbox{$\Lambda^0_{\mathrm{b}}$}}

\newcommand{\Zz}{\mbox{${\mathrm{Z}^0}$}}

\newcommand{\Jetset}{\mbox{JETSET}}

\newcommand {\ra}            {\mbox{$\rightarrow$}}

\begin{document}
\begin{titlepage}
 \begin{center}{\Large   EUROPEAN LABORATORY FOR PARTICLE PHYSICS
 }\end{center}\bigskip
\begin{flushright}
  CERN-EP/98-119 \\  
  20$^{\mathrm{th}}$ July 1998
\end{flushright}
%
\bigskip\bigskip\bigskip\bigskip\bigskip
\begin{center}
{\Huge \bf
Measurement of the Average} \\
\vspace{3mm}
{\Huge \bf Polarization of 
 b Baryons} \\
\vspace{1mm}
{\Huge \bf in Hadronic {\boldmath $\mathrm{Z}^0$\unboldmath} Decays}
\end{center}
\bigskip \bigskip 
\begin{center}{\LARGE The OPAL Collaboration}
\end{center}


\bigskip\bigskip

\begin{center}{\Large \bf Abstract}\end{center}

In the Standard Model, b quarks produced in 
$\mathrm{e}^+\mathrm{e}^-$ annihilation at the $\mathrm{Z}^0$ peak 
have a large average longitudinal polarization of 
$-0.94$. Some fraction of this polarization is expected to
be transferred to b-flavored baryons during hadronization.
The average longitudinal polarization
of weakly decaying b baryons, \plb,
is measured in approximately 4.3 million hadronic
$\mathrm{Z}^0$ decays collected with the OPAL detector
between 1990 and 1995 at LEP. Those b baryons that
decay semileptonically and produce a $\Lambda$ baryon are
identified through the correlation of the baryon number of the
$\Lambda$ and the electric charge of the lepton.
In this semileptonic decay, the ratio of the neutrino energy
to the lepton energy is a sensitive polarization observable. 
The neutrino energy is estimated using missing energy 
measurements.  From a fit to the distribution of this ratio,
the value
$$
\plb  = -0.56^{+0.20}_{-0.13} \pm 0.09
$$
is obtained, where the first error is statistical and the second systematic.

\vspace{0.25cm}
\begin{center}{
Submitted to {\sl Physics Letters B}
}
\end{center}


\end{titlepage}

\begin{center}{\Large        The OPAL Collaboration
}\end{center}\bigskip
\begin{center}{
G.\thinspace Abbiendi$^{  2}$,
K.\thinspace Ackerstaff$^{  8}$,
G.\thinspace Alexander$^{ 23}$,
J.\thinspace Allison$^{ 16}$,
N.\thinspace Altekamp$^{  5}$,
K.J.\thinspace Anderson$^{  9}$,
S.\thinspace Anderson$^{ 12}$,
S.\thinspace Arcelli$^{ 17}$,
S.\thinspace Asai$^{ 24}$,
S.F.\thinspace Ashby$^{  1}$
% ...
}\end{center}\bigskip
\bigskip
$^{  1}$School of Physics and Astronomy, University of Birmingham,
Birmingham B15 2TT, UK
\newline
$^{  2}$Dipartimento di Fisica dell' Universit\`a di Bologna and INFN,
I-40126 Bologna, Italy
\newline
$^{  3}$Physikalisches Institut, Universit\"at Bonn,
D-53115 Bonn, Germany
\newline
$^{  4}$Department of Physics, University of California,
Riverside CA 92521, USA
\newline
$^{  5}$Cavendish Laboratory, Cambridge CB3 0HE, UK
\newline
$^{  6}$Ottawa-Carleton Institute for Physics,
Department of Physics, Carleton University,
Ottawa, Ontario K1S 5B6, Canada
\newline
$^{  7}$Centre for Research in Particle Physics,
Carleton University, Ottawa, Ontario K1S 5B6, Canada
\newline
$^{  8}$CERN, European Organisation for Particle Physics,
CH-1211 Geneva 23, Switzerland
\newline
$^{  9}$Enrico Fermi Institute and Department of Physics,
University of Chicago, Chicago IL 60637, USA
\newline
$^{ 10}$Fakult\"at f\"ur Physik, Albert Ludwigs Universit\"at,
D-79104 Freiburg, Germany
\newline
$^{ 11}$Physikalisches Institut, Universit\"at
Heidelberg, D-69120 Heidelberg, Germany
\newline
$^{ 12}$Indiana University, Department of Physics,
Swain Hall West 117, Bloomington IN 47405, USA
\newline
$^{ 13}$Queen Mary and Westfield College, University of London,
London E1 4NS, UK
\newline
$^{ 14}$Technische Hochschule Aachen, III Physikalisches Institut,
Sommerfeldstrasse 26-28, D-52056 Aachen, Germany
\newline
$^{ 15}$University College London, London WC1E 6BT, UK
\newline
$^{ 16}$Department of Physics, Schuster Laboratory, The University,
Manchester M13 9PL, UK
\newline
$^{ 17}$Department of Physics, University of Maryland,
College Park, MD 20742, USA
\newline
$^{ 18}$Laboratoire de Physique Nucl\'eaire, Universit\'e de Montr\'eal,
Montr\'eal, Quebec H3C 3J7, Canada
\newline
$^{ 19}$University of Oregon, Department of Physics, Eugene
OR 97403, USA
\newline
$^{ 20}$CLRC Rutherford Appleton Laboratory, Chilton,
Didcot, Oxfordshire OX11 0QX, UK
\newline
$^{ 22}$Department of Physics, Technion-Israel Institute of
Technology, Haifa 32000, Israel
\newline
$^{ 23}$Department of Physics and Astronomy, Tel Aviv University,
Tel Aviv 69978, Israel
\newline
$^{ 24}$International Centre for Elementary Particle Physics and
Department of Physics, University of Tokyo, Tokyo 113, and
Kobe University, Kobe 657, Japan
\newline
$^{ 25}$Institute of Physical and Environmental Sciences,
Brunel University, Uxbridge, Middlesex UB8 3PH, UK
\newline
$^{ 26}$Particle Physics Department, Weizmann Institute of Science,
Rehovot 76100, Israel
\newline
$^{ 27}$Universit\"at Hamburg/DESY, II Institut f\"ur Experimental
Physik, Notkestrasse 85, D-22607 Hamburg, Germany
\newline
$^{ 28}$University of Victoria, Department of Physics, P O Box 3055,
Victoria BC V8W 3P6, Canada
\newline
$^{ 29}$University of British Columbia, Department of Physics,
Vancouver BC V6T 1Z1, Canada
\newline
$^{ 30}$University of Alberta,  Department of Physics,
Edmonton AB T6G 2J1, Canada
\newline
$^{ 31}$Duke University, Dept of Physics,
Durham, NC 27708-0305, USA
\newline
$^{ 32}$Research Institute for Particle and Nuclear Physics,
H-1525 Budapest, P O  Box 49, Hungary
\newline
$^{ 33}$Institute of Nuclear Research,
H-4001 Debrecen, P O  Box 51, Hungary
\newline
$^{ 34}$Ludwigs-Maximilians-Universit\"at M\"unchen,
Sektion Physik, Am Coulombwall 1, D-85748 Garching, Germany
\newline
\bigskip\newline
$^{  a}$ and at TRIUMF, Vancouver, Canada V6T 2A3
\newline
$^{  b}$ and Royal Society University Research Fellow
\newline
$^{  c}$ and Institute of Nuclear Research, Debrecen, Hungary
\newline
$^{  d}$ and Department of Experimental Physics, Lajos Kossuth
University, Debrecen, Hungary
\newline
$^{  e}$ on leave of absence from the University of Freiburg
\newpage


\section{Introduction}


\indent
According to the Standard Model, the process
$\mathrm{e^+ e^-} \rightarrow \mathrm{Z}^0 \rightarrow \mathrm{b\bar{b}}$
gives rise to b quarks that are longitudinally
polarized~\cite{Gphys,ManelSchuler92} with a large average value of  
$\langle P_L^{\mathrm{b}} \rangle = -0.94$ for
a weak mixing angle of 
$\sin^2\theta_W = 0.23$.  This polarization varies by only
$\pm 2\%$ over the full range of the production 
angle\footnote{In the OPAL coordinate system the $x$ axis
points towards the center of the LEP ring, the $y$ axis points
upwards and the $z$ axis points in the direction of the electron beam.
The polar angle $\theta$, the azimuthal angle $\phi$ and the radius
$r$ denote the usual spherical coordinates.} $\theta$.
If subsequent hadronization to a \lbz\
baryon is considered for example, 
the light u and d quarks form a spin-0 system and the 
spin of the \lbz\ should be carried entirely 
by the b quark.  In the heavy-quark limit, an important
prediction of Heavy Quark Effective Theory (HQET) is that
the degrees of freedom of the b quark are decoupled from the
spin-0 light diquark so that
the \lbz\ should retain almost 100\% of this 
polarization~\cite{ManelSchuler92} with only a slight reduction
of about 3\% due to hard gluon emission during
hadronization~\cite{gluons}.
This can be contrasted with spin-0 pseudoscalar B mesons 
where the polarization 
information of the b quark is lost
or vector B$^*$ meson where the 
polarization 
information is not expected to be observable~\cite{bmeson}.
In b baryon production, it is also possible for the b quark to
combine with a spin-1 uu, ud, or dd diquark system to form the
higher mass baryonic states $\Sigma_{\mathrm{b}}$ and
$\Sigma_{\mathrm{b}}^*$ that are expected to decay strongly to
$\lbz \pi$ leading to substantial reduction in the polarization
averaged over
the
weakly decaying b baryons~\cite{bmeson,korner}.  
Measuring the \lb\ polarization\footnote{The symbol \lb\ will
be used in this paper to refer to all b baryons that decay
weakly, including for example the \lbz\ and the b quark in combination
with a spin-1 us or ds diquark forming a \xb.
The symbol \lbz\ is used to denote
the particular ground state b baryon with quark content (bud).} therefore
provides a test of HQET and information about heavy baryon 
hadronization and nonperturbative corrections to spin transfer
in fragmentation. 

The sign of measured polarization
also gives information on the chirality of the b quark
coupling to the weak charged current~\cite{righthand}, particularly 
interesting because of recent analyses testing for the presence
of sizeable right-handed components in the 
$\mathrm{b} \rightarrow \mathrm{c}$ charged current 
coupling~\cite{righthints}.
However, in this case, a polarization values
of specific b baryon states would have to be assumed
to extract limits on the right-handed component of the charged current
and hence the limits would be strongly model dependent.

In the weak semileptonic decay 
$\lb \rightarrow \mathrm{X_c} \ell^- \bar{\nu}_{\ell} \mathrm{X}$, both
the charged lepton and neutrino energy spectrum are 
sensitive to \plb~\cite{Gphys,ManelSchuler92,Mele}.
Assuming that b decays proceed via the usual left-handed 
current
(i.e., $(V-A)$ coupling), 
in the rest frame of the \lb, 
charged leptons $\ell^-$ tend to be emitted
antiparallel to the spin of the \lb\ and the 
$\bar{\nu}_{\ell}$'s  parallel to
the \lb\ spin.  In the laboratory frame, polarization then implies
a harder lepton energy spectrum and softer neutrino energy
spectrum compared to the unpolarized case.
However, there are substantial uncertainties in the
exact shape of these spectra due to
uncertainties in fragmentation,
the ratio of the quark masses $m_{\mathrm{c}}/m_{\mathrm{b}}$,
and QCD corrections~\cite{qcd1,qcd2} such that the extraction
of the average polarization from the spectra alone is problematic.
These systematic effects partially
cancel in the ratio of the average of the
lepton energy to the  average of the neutrino energy or 
in the ratios of higher moments of the energy 
spectra~\cite{ratios1,ratios2}.

This paper describes a measurement
of the average longitudinal polarization of weakly decaying b baryons,
\plb, using about 4.3 million multihadronic \Zz\ decays collected by the
OPAL detector from 1990 to 1995.
To ensure a data sample with reasonably large statistics,
events containing
decay chains of the form\footnote{Charge conjugate processes 
are implied throughout this paper.}
$\lb  \rightarrow \Lambda^+_{\mathrm{c}} 
\ell^- \bar{\nu}_{\ell} \mathrm{X}$ followed by 
$\Lambda^+_{\mathrm{c}} \rightarrow \Lambda \mathrm{X}$
are selected without reconstructing the intermediate charm state.
The correlation of a $\Lambda$ with a negatively-charged
lepton ($\ell = \mathrm{e}$ or $\mu$) 
or a $\bar{\Lambda}$ with a positively-charged lepton can
indicate the presence of a semileptonic b-baryon decay, as
used previously by OPAL~\cite{lbcorold,lbcorrel}.
The other charge combinations ($\bar{\Lambda}$-$\ell^-$ and
$\Lambda$-$\ell^+$) are used to characterize
the background.
The neutrino energy $E_{\nu}$ was estimated
using missing energy measurements in the hemisphere containing the
$\Lambda$-$\ell^-$ pair.
To extract the polarization of the b baryon,
a fit was then made to the distribution of the ratio
$E_{\nu}/E_{\ell}$ formed event by event. 
Fitting to the
shape of this distribution should lead to an estimate
with lower statistical variance than the use of
the average sample energies 
$\langle E_{\nu} \rangle / \langle E_{\ell} \rangle$ as employed in
the previously published
measurement of \plb~\cite{alephpub}.


\section{The OPAL detector}

\indent
A detailed description of the OPAL detector can be found 
elsewhere~\cite{bib-OPALdet}.  
The almost complete hermeticity of the OPAL detector allows
the missing energy in an event to be determined effectively.
The central tracking detector consists of a two-layer silicon
microstrip detector  with polar 
angle
coverage $|\cos\theta| < 0.8$ immediately surrounding
the beam-pipe, 
followed by a high-precision vertex drift chamber,
a large-volume jet chamber
that records the momentum and energy loss of charged particles
over 98\% of the solid angle,
and chambers which 
measure the $z$-coordinate, all in a uniform
0.435~T axial magnetic field.  
Charged particles can be identified by their specific ionization
energy loss, d$E$/d$x$, in the jet chamber.  
Further information
on the performance of the tracking and d$E$/d$x$ measurements can
be found in Ref.~\cite{tracking}.
A lead-glass electromagnetic calorimeter
is located outside the magnet coil, which, in combination with the 
forward calorimeter, gamma catcher and silicon-tungsten 
luminometer~\cite{SW}, complete the geometrical acceptance down 
to 24~mrad from the beam direction.  
The magnet return yoke is instrumented with streamer tubes for hadron
calorimetry and is surrounded by several layers of muon chambers.

\section{Monte Carlo simulations}

\indent
To model the signal, samples of simulated data were generated
of \Zz\ hadronic decays that included
$\lbz \rightarrow \Lambda^+_{\mathrm{c}}
\ell^- \bar{\nu}_{\ell} \mathrm{X}$ followed by 
$\Lambda^+_{\mathrm{c}} \rightarrow \Lambda \pi^+ \mathrm{X}$ and
$\Lambda \rightarrow \mathrm{p}\pi^-$
using a modified JETSET decay routine
incorporating polarized \lbz\ decay with QCD corrections as
described in Ref.~\cite{qcd2}.  Samples of 
$40 \thinspace 000$ events
each were generated at twelve different values of \plb.
These signal samples and the heavy flavor Monte Carlo event samples
described below
were generated using the latest world average branching ratios
for \lc\ decays~\cite{PDG}.
Additional samples including primary production and
decay of $\Sigma_{\mathrm{b}}$
or \xb\ baryons were used to assess the effect 
of the presence of different baryons on the result. 

Monte Carlo simulation samples of inclusive hadronic
$\mathrm{Z^0}$ decays were used to check backgrounds.
The JETSET 7.4 parton shower Monte Carlo generator~\cite{pythia}
with the fragmentation function of Peterson {\it et al.}~\cite{peterson}
for heavy quarks was used to generate samples including approximately
4~million hadronic $\mathrm{Z^0}$ decays and 2~million
$\mathrm{Z^0} \rightarrow \mathrm{b\bar{b}}$ decays
(the equivalent of about 9~million hadronic decays).
As will be discussed later, in the decay
$\mathrm{b} \rightarrow \tau \overline{\nu}_{\tau} \mathrm{X}$
when the $\tau$ decays leptonically, 
the resulting lepton and neutrino background has to be taken into account
carefully.
The tau polarization in these decays was calculated
according to Ref.~\cite{falktau} and leptonic decays
of these and all other $\tau$ leptons were handled by the
TAUOLA package~\cite{tauola}. 

All Monte Carlo samples were processed through a full
simulation of the OPAL detector~\cite{bib-GOPAL} and the
same event analysis chain was applied to the simulated events
as to the data.

\section{Data and event selection}

\indent
This analysis uses data collected at center-of-mass energies within
$\pm 3$~GeV of the $\mathrm{Z^0}$ mass
during the 1990--1995 LEP 
running periods.
Following detector performance requirements, track quality cuts, and
a standard hadronic event selection~\cite{bib-hadsel},
a data sample of 4.3~million multihadronic events was selected.
Given the importance of having well contained events in the
detector for a reliable measurement of the missing energy, 
each event was required to have $|\cos\theta_{\mathrm{thr}}| < 0.85$
where $\theta_{\mathrm{thr}}$ is the polar angle of the thrust
axis.  

The selection of b-baryon decays using
$\Lambda$-lepton pairs is similar to the one previously used
to measure the average b-baryon lifetime and production
rate~\cite{lbcorrel} but 
uses improved electron identification, a different jet definition, 
different lepton kinematic cuts, and 
the inclusion of an additional $\pi^+$ to exploit the charge
correlation of $\Lambda$-$\pi^+$ in the decay chain
$\Lambda_{\mathrm{b}} \rightarrow \Lambda_{\mathrm{c}}^+
                      \rightarrow \Lambda \pi^+ {\mathrm{X}}$.
These all contribute to an overall improvement of the 
signal-to-purity ratio by 45\%.
Firstly, electrons and muons having large momentum and large transverse
momentum with respect to their associated jet (including the lepton
candidate) were identified. 
$\Lambda$ baryons were then reconstructed and combined with $\pi^+$'s
that were consistent with coming from a common vertex formed by the
$\Lambda$ and lepton candidate.  
In the signal process the $\pi^+$ most commonly comes from the
\lcp\ decay with a decay length from the point of \lb\ decay
which is small compared with the typical
$\Lambda$-lepton vertex resolution.
Requirements were placed on the 
$\Lambda$-lepton combination followed by  requirements 
to improve the missing energy determination and its applicability
in estimating the neutrino energy in the semileptonic decay of the \lb.

Electrons were identified
using an artificial neural network which
is a simplified version of that described in 
Ref.~\cite{elecid}, using only six rather than twelve inputs. 
These are: the momentum and polar angle of the track, the
energy-momentum ratio $E/p$, the number of electromagnetic calorimeter
blocks contributing to the energy measurement, the
ionization energy loss 
and its error.
Photon conversion candidates were rejected using another
neural network algorithm,
similar to that described in Ref.~\cite{elecid}, but using the new electron
identification algorithm to identify the conversion partner
to the electron candidate.
Muon candidates were identified by associating track segments in the
outer muon detectors with tracks extrapolated from the central tracking
detectors~\cite{muonid}.

Charged tracks and electromagnetic clusters
not associated with charged tracks were combined into jets
using a cone algorithm~\cite{jetcone} with
a minimum jet energy of 5.0~GeV and cone half angle of 550~mrad.  
Electron and muon candidates were required to have momentum greater
than 4.0~GeV and transverse momentum with respect to the associated
jet axis, $p_t$,
greater than 0.7~GeV to obtain a sample enriched in
b-quark events.

Candidate $\Lambda$ baryons decaying via
$\Lambda \rightarrow \mathrm{p} \pi^-$ were identified
as described in Ref.~\cite{lbcorrel} by considering oppositely charged
tracks assigned to the same jet as the lepton
and satisfying d$E$/d$x$ criteria and various kinematic and
geometric requirements.  To reduce 
$\mathrm{K}^0_S \rightarrow \pi^+ \pi^-$ contamination,
the invariant mass, 
measured assuming a pion mass
for both particles, was required to be
more than one standard
deviation in mass resolution away from the $\mathrm{K}^0_S$ mass:
less than 0.491~GeV or greater than 0.503~GeV.
To decrease the contribution of $\Lambda$ candidates
coming from fragmentation processes, the $\Lambda$ candidate
was required to have a minimum momentum of 4~GeV.
The invariant mass
of the reconstructed $\Lambda$ was required
to be between 1.1078 and 1.1234~GeV.
The fraction of $\Lambda$ baryons wrongly identified as
$\bar{\Lambda}$ is negligible since for these 
$\Lambda$ momenta the proton candidate
can always be taken as the higher momentum track.
The efficiency of this selection varies from 23\% to 10\%
for low to high momentum $\Lambda$ candidates.

Lepton and $\Lambda$ candidates were then correlated using the following
criteria.  The reconstructed momentum vector of the $\Lambda$ candidate
was required to be
closer than 50$^{\circ}$ to the momentum vector of the lepton candidate.
The invariant mass of the $\Lambda$-lepton system was demanded to
be greater than 2.2~GeV to reduce the background from  
$\lcp$'s that decay 
semileptonically: 
$\lcp \rightarrow \Lambda \ell^+ \nu_{\ell} \mathrm{X}$.
To reduce further random combinatorial background, the 
magnitude of the momentum vector
sum of the $\Lambda$ and lepton had to be greater than 9~GeV.

For each accepted $\Lambda$-lepton pair, candidate
$\pi^+$'s were selected among all charged tracks in the jet to which
the lepton was assigned.
Charged tracks with momentum greater than
0.5~GeV and with measured d$E$/d$x$ having a probability greater than
1\% of being consistent with a pion were considered 
for combining in the vertex.  
The track was
required to be positively charged if combined with a $\Lambda$, and
negatively charged if combined with a $\bar{\Lambda}$.
A decay vertex was formed in the $r$-$\phi$ plane
by extrapolating the candidate $\Lambda$
momentum vector from its decay vertex to the intersection with the 
lepton candidate track.
The $\pi$ candidate was then
also fitted to the $\Lambda$-lepton decay vertex and  
the candidate track
with the largest corresponding $\chi^2$ vertex probability 
was chosen.
$\Lambda \ell^- \pi^+$ combinations were retained if the $\chi^2$ probability
of the common fit vertex was greater than 1\%
and the invariant mass of the
common vertex less than 5.6~GeV.

Even for properly reconstructed
$\Lambda \ell$ combinations coming from the decay 
$\lb \rightarrow \lcp \ell^- \bar{\nu}_{\ell}$, the 
subsequent semileptonic decay of
$\lcp \rightarrow \Lambda \ell^+ \nu_{\ell}$ would result in a second neutrino
compromising the measurement of $E_{\nu}$ from the \lb\ decay using
the missing energy in the event.
Therefore a small fraction of the $\Lambda \ell$ combinations were rejected 
if the $\pi^+$ candidate track selected above 
had momentum greater than 2.0 GeV and was identified
as an electron or muon.

Each event was divided into two hemispheres by a plane 
perpendicular to the thrust axis.
The hemisphere containing the $\Lambda \ell^- \pi^+$ combination
was defined as the signal hemisphere and the opposite hemisphere
as the recoil hemisphere.  
To reject events with incorrectly measured 
total energy and therefore fake missing energy due to mismeasurement of 
momenta, signal
hemispheres were rejected if they contained high-momentum tracks close
to anode planes in the jet chamber, i.e., with measured momenta 
greater than 10~GeV and satisfying 
$|\phi - \phi_{\mathrm{anode}}| < 1^{\circ}$ where
$\phi$ is the azimuthal angle of the track and
$\phi_{\mathrm{anode}}$ is the angle of the closest 
anode wire plane in the jet chamber.
In the hemisphere opposite to the signal hemisphere, additional
missing energy can arise due to semileptonic decays of b and c
hadrons into electrons and muons.  
Signal hemispheres were rejected
if the opposite hemisphere contained an electron or muon satisfying
the kinematic cuts described earlier since the presence of a lepton
can indicate the possible presence of a neutrino with substantial 
energy in the opposite hemisphere from a semileptonic decay.

\begin{figure}[!htb]
\begin{center}
\epsfxsize=10.0cm
\epsffile{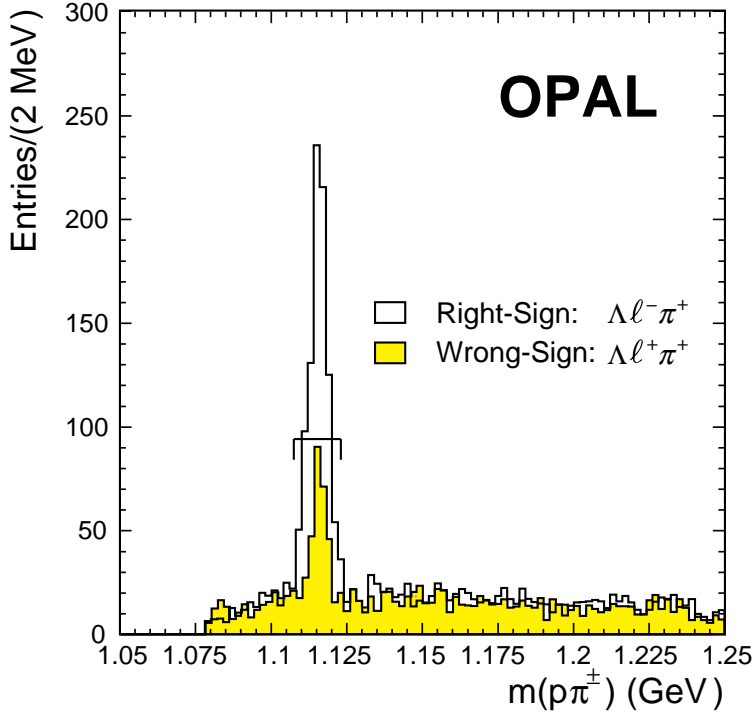}
\caption[]{{\sl
Invariant mass distribution of
${\mathrm{p}}\pi^-$ combinations in the right-sign
$\Lambda \ell^- \pi^+$ sample (open histogram) superimposed by
wrong-sign $\Lambda \ell^+ \pi^+$ (shaded histogram) combinations.
Charge conjugate states are implied if not stated otherwise.
The signal region is shown by the bracketed range of masses.
The peak at the $\Lambda$ mass in the wrong-sign distribution is
indicative of genuine $\Lambda$ particles (though not necessarily from
b-baryon decay, e.g.\ from fragmentation) being combined with real
or fake leptons (i.e., hadrons misidentified as leptons).}}
\label{fig:signal}
\end{center}
\end{figure}

The invariant mass distribution of the
$\mathrm{p} \pi^-$ combination is shown in Fig.~\ref{fig:signal} for
$\Lambda \ell^- \pi^+$ (right-sign) and $\Lambda \ell^+ \pi^+$ (wrong-sign)
combinations after all other  selection requirements have been applied.
A total of 912 right-sign and 316 wrong-sign
$\Lambda \ell^- \pi^+$ combinations are selected with an overall
b baryon purity of 69\% in the right-sign sample.
The 596 excess right-sign combinations can be attributed mainly
to b-baryon decays. 
The overall efficiency to select true $\Lambda$'s and leptons
from b baryon semileptonic decays is estimated from Monte Carlo
simulated samples to be approximately 5.6\%,
although a 
knowledge of this efficiency is not necessary for the polarization
measurement.
The number of observed wrong-sign combinations will be used
after small corrections
as an estimate of the level of the background as well as to
estimate the shape of the background in the distribution of the
ratio $E_{\nu}/E_{\ell}$.

\section{Neutrino and lepton energy measurement}

\indent
In the semileptonic decay of the \lb, the neutrino energy was
estimated by the measured missing energy in the signal hemisphere, which
can be defined by
$$
E_{\nu} = E^{\mathrm{hemi}}_{\mathrm{miss}} = 
E_{\mathrm{beam}} + E_{\mathrm{corr}} - 
E^{\mathrm{hemi}}_{\mathrm{vis}},
$$ 
where $E_{\mathrm{beam}}$ is the beam energy and 
$E^{\mathrm{hemi}}_{\mathrm{vis}}$ the visible energy in the
signal hemisphere.  If we assume a two-body decay of the 
$\mathrm{Z}^0$, with each hemisphere considered as one body, then
$ E_{\mathrm{corr}}$ is a correction term
calculated assuming that the two hemisphere momenta are equal: 
$$
 E_{\mathrm{corr}} = (M^2_{\mathrm{sig}} - M^2_{\mathrm{recoil}})/
4 E_{\mathrm{beam}}.
$$
$M_{\mathrm{sig}}$ and $M_{\mathrm{recoil}}$ are the measured invariant
masses of the signal hemisphere and opposite recoil hemisphere, respectively.
The term $E_{\mathrm{corr}}$ 
uses the beam energy constraint to improve the resolution for the missing
energy in the signal hemisphere by accounting  for fluctuations of
energy splitting 
between hemispheres 
and unobserved neutral energy
on an event by event basis.

Calculations of these quantities used the four-momenta of reconstructed
charged tracks and of clusters in the electromagnetic and hadronic
calorimeters not associated with charged tracks.  The masses of all
charged particles were set to the charged pion mass except
for the proton from the identified $\Lambda$ and the identified
lepton which are set to their appropriate masses. The invariant
masses of the calorimeter energy clusters were assumed to be zero.
Calorimeter clusters associated with charged tracks were also included
after the expected calorimeter energy for the associated track
is subtracted from the cluster energy to reduce double 
counting~\cite{bib-MT}.  
If the energy of a cluster was smaller than the expected energy for
the associated tracks, the cluster energy is not used.

\begin{figure}[!htb]
\begin{center}
\epsfxsize=15.0cm
\epsffile{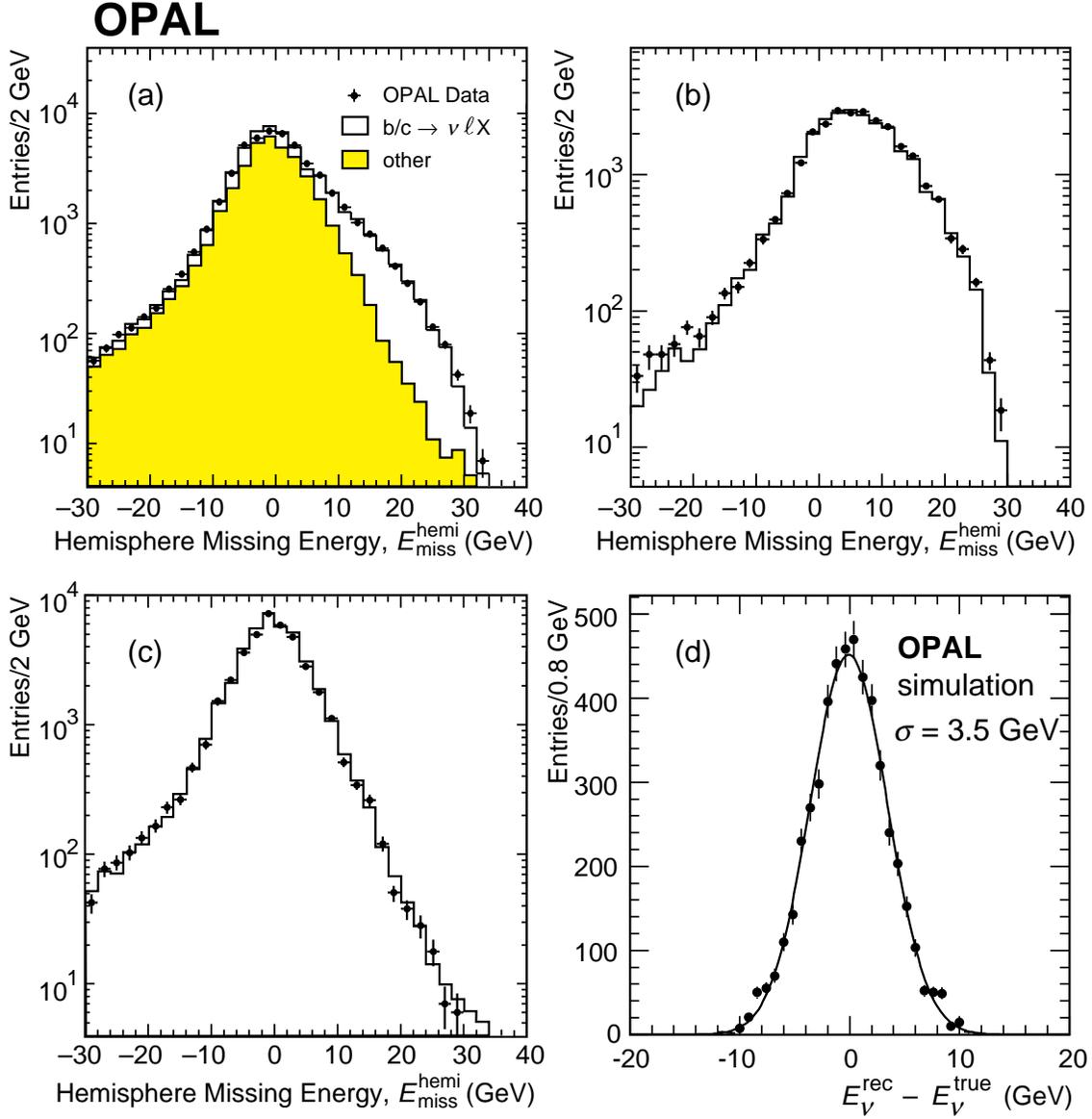}
\caption[]{{\sl
(a) Reconstructed missing energy in data  
(points with error bars)
and Monte Carlo simulated events in ${q\bar{q}}$
events after a preselection with relaxed $p$ and $p_t$
requirements; the open histogram
shows the missing energy for events containing b and c hadrons decaying 
semileptonically,
and the shaded histogram is for all other processes.
Reconstructed missing energy for tagged $b\bar{b}$ events
in data (points with error bars)
and Monte Carlo simulated events (open histograms)
(b) with an identified lepton in the
hemisphere opposite the b-tag and (c) with no identified lepton in the
hemisphere opposite the b-tag.
(d) Neutrino energy resolution from reconstructed $E_{\nu}^{\mathrm{rec}}$
and true $E_{\nu}^{\mathrm{true}}$ in Monte Carlo signal events with
$\lbz \rightarrow \Lambda^+_{\mathrm{c}}
\ell^- \bar{\nu}_{\ell} \mathrm{X}$.
}}
\label{fig:emiss}
\end{center}
\end{figure}

The agreement of the quantity $E^{\mathrm{hemi}}_{\mathrm{miss}}$
between data and Monte Carlo events 
containing identified leptons following
a preselection with relaxed $p$ and $p_t$ cuts to enhance statistics
is shown in Fig.~\ref{fig:emiss}(a).
To further test and calibrate the measurement of this important
quantity, event samples from both data and Monte Carlo simulation
were prepared that were enhanced in b quarks
using a lifetime tag in one hemisphere.
Hemispheres were identified as containing b hadrons
by reconstructing
secondary vertices and using the parameters
of the vertices in an artificial neural network
trained to reject non-b background~\cite{btag}.
The most important inputs to the neural network were
the decay length in the $x$-$y$ plane, its uncertainty, 
and the number of tracks in the vertex.
The selected sample had a b-quark purity of approximately
90\%. 

Two control samples were used: one with an electron or muon candidate
identified as previously described in the hemisphere opposite the
b-tagged hemisphere and another with no lepton identified in the
hemisphere opposite the b-tagged hemisphere.  The first sample
is enriched in energetic neutrinos while the second sample
is depleted. 
From the distributions of the measured and true value of
$E^{\mathrm{hemi}}_{\mathrm{miss}}$ in these Monte Carlo samples, 
small residual neutrino energy shifts were observed.  
These could be approximated well with
an additional average 
linear correction, determined from both control samples,
subtracted from the reconstructed value of 
$E^{\mathrm{hemi}}_{\mathrm{miss}}$ that ranged from 50~MeV
for $E_{\nu}^{\mathrm{true}} = 0$~GeV (i.e., no neutrino) to 910~MeV for 
$E_{\nu}^{\mathrm{true}} = 30$~GeV.
This correction was applied to both the data and Monte Carlo samples
to eliminate the small bias inherent in the neutrino energy reconstruction
method.
The agreement between data and Monte Carlo 
simulated events in these control samples
can be seen in Fig.~\ref{fig:emiss}(b) and (c).
Independent samples of events with hemispheres failing the
b-tag requirement, both with and without identified leptons, also
show good agreement between data and Monte Carlo.

The absolute resolution on the missing energy and hence the neutrino energy
improves with increasing neutrino energy from $\sigma(E_{\nu}) = 5.0$~GeV
at $E_{\nu} = 0$~GeV to 2.9~GeV at  $E_{\nu} = 15$~GeV.
In the signal Monte Carlo samples,
the average neutrino energy resolution was observed to be 
3.5~GeV as shown in 
Fig.~\ref{fig:emiss}(d).  The average neutrino energy for the 
unpolarized signal
process is about 6.1~GeV with an r.m.s.\ spread of 4.4~GeV after 
applying only
kinematic and geometric requirements but not including detector resolution 
effects.

For the electron and muon candidates
the momentum was used as an estimate of the lepton energy.
For high energy electrons, 
the electromagnetic cluster energy
associated with the electron may have better resolution,
but the momentum was used to avoid biases due to overlapping energy
deposits of nearby particles.  Changes in the electron spectrum
due to final-state radiation are included in the Monte Carlo
simulation.
The distributions of lepton spectra in the control samples described
above were also compared between data and Monte Carlo.  The shapes
agree well and the average values are 
$\langle E_{\ell} \rangle = 10.18 \pm 0.03$~GeV for the data and 
$\langle E_{\ell} \rangle = 10.11 \pm 0.03$~GeV for the Monte Carlo simulation
for the b-tagged sample containing identified leptons.

\begin{figure}[!htb]
\begin{center}
\epsfxsize=12.0cm
\epsffile{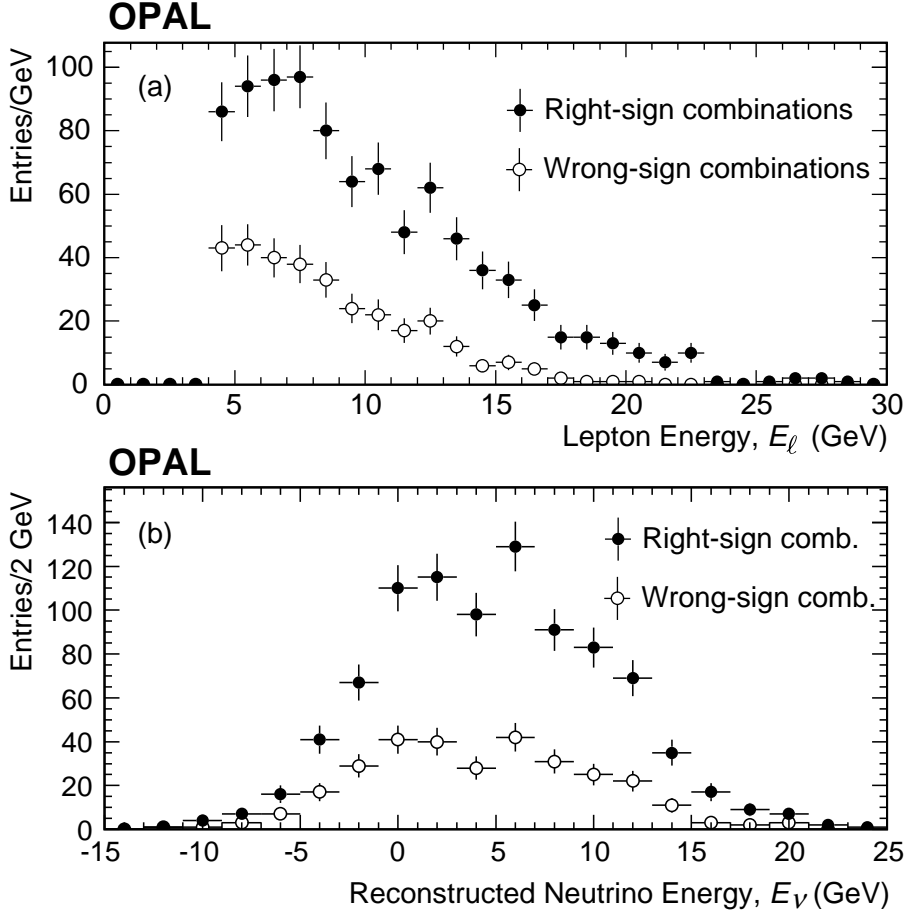}
\caption[]{{\sl
(a) Lepton (electron and muon) and (b) reconstructed neutrino
energy spectra for the
hemispheres containing selected right-sign $\Lambda \ell^- \pi^+$
combinations
with the spectra for wrong-sign combinations overlaid.}}
\label{fig:spectra}
\end{center}
\end{figure}

The distributions of the measured electron and muon energies and of the
reconstructed neutrino energies for the selected right-sign
$\Lambda \ell^- \pi^+$ and wrong-sign $\Lambda \ell^+ \pi^+$ 
samples in the data are
shown in Fig.~\ref{fig:spectra}.

\section{Background estimate}

\indent
The wrong-sign sample 
is used as an estimate of the level and
shape of the background in the right-sign sample.
The component of the background due to random combinations of tracks and
fakes due to misidentified leptons or $\Lambda$ baryons is expected
to contribute equally to right- and wrong-sign samples, as verified
by Monte Carlo simulations.  

A larger source of background is the
combination of a $\Lambda$ baryon from the fragmentation process
with a genuine lepton from the semileptonic decay of b or c hadrons.
In the framework of string fragmentation, there is a strong correlation
between, for example, the \lb\ subsequently decaying as
$\lb \rightarrow \lcp \ell^- \bar{\nu}_{\ell} \mathrm{X}$
and a high momentum $\bar{\Lambda}$ as one of the neighbouring particles
in the fragmentation chain.  In this case, the background of a fragmentation
$\bar{\Lambda} \ell^-$ combination will preferentially enter the
wrong-sign sample.  Table~\ref{tab:back} gives the
background composition of the right- and wrong-sign sample as predicted
by Monte Carlo simulation and shows the resulting imbalances.
These predictions use the ``popcorn'' model~\cite{popcorn}
and the
indicated uncertainties include modelling effects as described in 
detail in Ref.~\cite{lbcorrel}.
Note that even if a true lepton from a semileptonic
\lb\ decay is combined with a fragmentation $\Lambda$, 
the ratio $E_{\nu}/E_{\ell}$ is
still indicative of b-baryon polarization
since missing energy and hence the neutrino energy is
calculated using event quantities. 
Events of this type are therefore treated as signal in the fitting procedure
(see below).

 \begin{table}[!htb]
  \begin{center}
    \begin{tabular}{| rl | r@{ }c@{ }l | r@{ }c@{ }l |}
      \hline
        \multicolumn{2}{| c |}{Source} & 
        \multicolumn{3}{ c |}{RS}   &
        \multicolumn{3}{ c |}{WS}  \\
      \hline \hline
      \multicolumn{2}{| l |}{Signal: $\Lambda$ and $\ell$ from b baryon} 
                    &  
        629 & $\pm$ &  30 & \multicolumn{3}{ c |}{ }    \\ \hline
      \multicolumn{2}{| l |}{Fragmentation $\Lambda$ plus:} 
                    &  
        \multicolumn{3}{ c |}{ } & \multicolumn{3}{ c |}{ }    \\
          & $\ell$ from B~meson 
                    & \hfill  94 &$\pm$& 9 \hfill & 
                       77 & $\pm$ & 8   \\
          & $\ell$ from c~baryon 
                    &    9 & $\pm$ & 3  & 
                         3 & $\pm$ & 2    \\
          & $\ell$ from b~baryon 
                    &   5 & $\pm$ & 4  & 
                        66 & $\pm$ & 6    \\
          & $\ell$ from c~meson
                    & 32 & $\pm$ & 6  &        
                      43 & $\pm$ & 7    \\
      \hline
      \multicolumn{2}{| l |}{Exclusive Backgrounds:} 
                    & 32 & $\pm$ & 7  &  27 & $\pm$ & 6   \\
      \hline
      \multicolumn{2}{| l |}{Combinatorials and Fakes:} 
                    & 111 & $\pm$ & 8  & 100 & $\pm$ & 8  \\       
       \hline     
    \end{tabular}
  \end{center}
  \caption{\label{tab:back} \sl
   Monte Carlo predictions of the composition among the 912 right-sign (RS)
combinations and the 316 wrong-sign combinations.
Errors are due to Monte Carlo statistics, branching ratio
uncertainties,  and modelling systematic 
errors (see Ref.~\cite{lbcorrel}).
  \normalsize }
\end{table}

A number of exclusive channels as described below
are expected to contribute to 
the right-sign and wrong-sign samples, but the total contributions
are predicted to be small compared to the background due to combinatorials
and fakes and fragmentation $\Lambda$ baryons.
For example, a b baryon can decay semileptonically into a $\tau$ lepton that
subsequently decays
into an electron and muon to enter into the right-sign sample as
background.  The $\tau$ polarization in this case was varied within
its range of uncertainties~\cite{falktau} allowing the \lb\ polarization to
vary between 0 and $-1$ to estimate the systematic error on the 
Monte Carlo prediction for this rate.  
The expected contribution
to the wrong-sign sample for leptons arising from
the process $\lcp \to \Lambda \ell^+ \nu_{\ell}$ is kept small by 
the cut on the $\Lambda \ell$ invariant mass.
The uncertainty in this rate can be affected by uncertainties in the
polarization
of the \lc\ which is varied by reweighting Monte Carlo simulated events
as described in Section 9.
Other exclusive backgrounds include small fractions of
$\lb \to \lcp \mathrm{D}^-_{\mathrm{s}}$ followed by
$\mathrm{D}^-_{\mathrm{s}} \rightarrow  \ell^- \mathrm{X}$ and
$\bar{\mathrm{B}} \to \mathrm{Y}_{\mathrm{c}} \bar{\mathrm{N}} \ell \bar{\nu}$,
(where $\mathrm{Y_c}$ denotes any c baryon)~\cite{lbcorrel}.

\section{Fitting procedure}

\indent
The b-baryon polarization was extracted by comparing the 
reconstructed distribution of $E_{\nu}/E_{\ell}$ in the data
to spectra estimated from fully simulated Monte Carlo events
corresponding to various values of \plb\ and satisfying all the
selection criteria.
A binned maximum likelihood fit was used to extract \plb\
by determining which Monte Carlo spectrum gives the best
description of the data.
To obtain the Monte Carlo spectrum for arbitrary values of
\plb, polynomial fits are made bin by bin as a function of \plb\
to allow interpolation between spectra. 
The binned likelihood method has several advantages
over analytical functions used to describe the data distributions
and resolutions in an unbinned likelihood fit: 
it takes into account correlations
between $E_{\nu}$, $E_{\ell}$, $\sigma(E_{\nu})$ and $\sigma(E_{\ell})$;
effects of kinematic cuts such as the minimum lepton energy;
energy, momentum, and missing energy resolution with
a full simulation of the detector response; and 
any variation of the selection efficiency with $E_{\nu}/E_{\ell}$.

The probability density function (PDF) was estimated by the binned
and normalized Monte Carlo spectrum with 
${\mathcal{P}}_i^{\mathrm{sig}}(\plbn)$
being the value of the PDF in bin $i$ of 30 bins for the 
predicted signal distribution for the average polarization value \plb. 
${\mathcal{P}}_i^{\mathrm{back}}$ is the normalized Monte Carlo spectrum
of the wrong-sign backgrounds listed in Table~\ref{tab:back} 
excluding the contribution of a fragmentation $\Lambda$ combined
with a genuine lepton from a b-baryon decay.
The expected number of entries in bin $i$ of the right-sign and
wrong-sign distributions are $n_i^{\mathrm{RS}}$ and
$n_i^{\mathrm{WS}}$ given by:
\begin{eqnarray*}
     n_i^{\mathrm{WS}} & = 
  & {\mathcal{P}}_i^{\mathrm{WS}}(\plbn) \cdot 
  N_{\mathrm{tot}}^{\mathrm{WS}} \\
     n_i^{\mathrm{RS}} & = 
  & \left[ (1 - f^{\mathrm{WS}}) {\mathcal{P}}_i^{\mathrm{sig}}(\plbn) +
     f^{\mathrm{WS}} {\mathcal{P}}_i^{\mathrm{WS}}(\plbn) \right]
             \cdot  N_{\mathrm{tot}}^{\mathrm{RS}}, \\
{\mathrm{with}} \thinspace \thinspace \thinspace
{\mathcal{P}}_i^{\mathrm{WS}}(\plbn) & = 
  & (1 - f^{\mathrm{WS}}_{\mathrm{sig}}) {\mathcal{P}}_i^{\mathrm{back}}
   + f^{\mathrm{WS}}_{\mathrm{sig}}{\mathcal{P}}_i^{\mathrm{sig}}(\plbn). 
\end{eqnarray*}
Here $f^{\mathrm{WS}}_{\mathrm{sig}}$ is the estimated fraction of
fragmentation $\Lambda$ combined with leptons from b baryon decays
as given in Table~\ref{tab:back},  $N_{\mathrm{tot}}^{\mathrm{RS}}$ and
$N_{\mathrm{tot}}^{\mathrm{WS}}$ are the total number of right-sign and
wrong-sign combinations observed in the data, and
$f^{\mathrm{WS}} = 
N_{\mathrm{tot}}^{\mathrm{WS}}/N_{\mathrm{tot}}^{\mathrm{RS}}$.

The binned likelihood $\mathcal{L}$ is  the product of 
Poisson probabilities for obtaining the numbers of events
observed in the data in each bin using the Monte Carlo expectation in 
that bin
for a given \plb\ for both the wrong-sign and right-sign spectra.
A normalization constraint was imposed on the total number of
observed right-sign combinations.
An adjusted likelihood~\cite{barlowstat} was used to take into account
statistical fluctuations in the Monte Carlo prediction.
The above assumes that both the level and shape of the
wrong-sign sample can be used to estimate the background in 
the right-sign sample.  After removing the signal component in the
Monte Carlo wrong-sign combinations, the distribution of 
$E_{\nu}/E_{\ell}$ is consistent with the distribution of the
background in the right-sign Monte Carlo combinations.  
Effects of differences are addressed as systematic errors below.

Further checks of the fitting procedure were performed using Monte Carlo
samples. Fully simulated Monte Carlo subsamples of known \plb, each
corresponding to the same number of combinations observed in the
data were used as input to the fit and no bias was observed at the
precision studied.  A simple Monte Carlo program was used to
generate an ensemble of 2000 samples, each with the same statistics
as selected in the data.  For a given value of \plb,
lepton and neutrino energies were 
randomly sampled
from analytical joint energy distributions, 
smeared according to parameterized
energy resolution functions, and kinematic cuts applied.  
The background and distribution was generated assuming
zero polarization and the wrong-sign distribution sampled
separately, including a signal component as expected in the data.
In all cases, no significant additional systematic biases
were observed, and residuals divided by the error on \plb\
indicated a correct evaluation of statistical errors by the likelihood
fit.  
Lastly, the complete analysis was applied directly to the 
approximately 4~million fully
simulated hadronic Monte Carlo events.
The fitted value of $\plb = -0.09^{+0.17}_{-0.13}$ is consistent
with a value of zero polarization for b baryons 
as simulated in the Monte Carlo sample.

\section{Fit result}

\begin{figure}[!htb]
\begin{center}
\epsfxsize=10.0cm
\epsffile{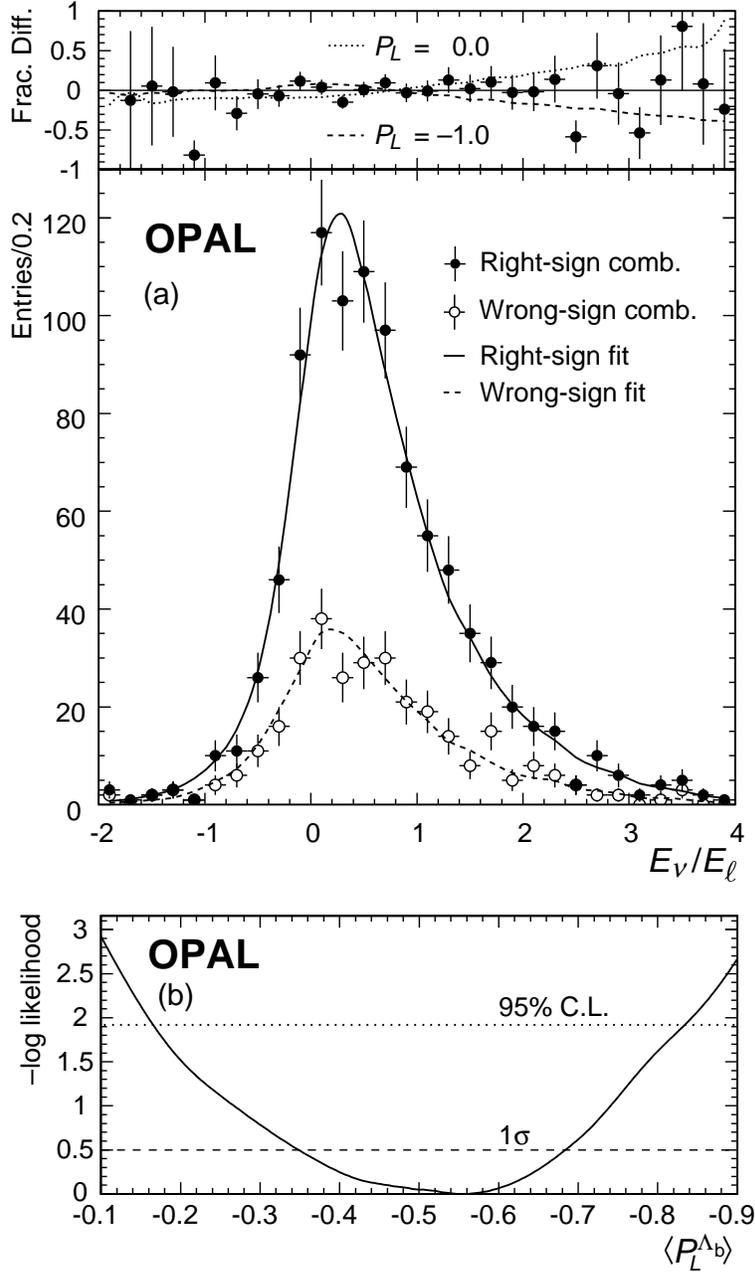}
\caption[]{{\sl
(a) Bottom: distribution of the reconstructed neutrino energy $E_{\nu}$ 
divided by the electron or muon energy $E_{\ell}$ for each
right-sign $\Lambda \ell^- \pi^+$ combination in the data (solid circles)
overlaid by the distribution for each of the wrong-sign combinations
(open circles). 
The solid and dashed lines show the result of the likelihood
fit to the right-sign and wrong-sign distribution, respectively.
Top: fractional difference 
$(n_i^{\mathrm{RS,data}} - n_i^{\mathrm{RS,fit}})/n_i^{\mathrm{RS,fit}}$
(points with error bars) compared to the analogous distributions
for the prediction for $\plb = 0.0$ (dotted line) and
$\plb = -1.0$ (dashed line).
(b) Negative log likelihood as a function of \plb\ for the
fit.  The dashed and dotted lines indicate the likelihood
change required for one standard deviation errors and 95\% CL
limits, respectively (statistical errors only).
}}
\label{fig:result}
\end{center}
\end{figure}

\indent
The 
event-by-event distribution of $E_{\nu}/E_{\ell}$ was
formed, as shown in Fig.~\ref{fig:result}(a), and the right-sign
and wrong-sign distributions were fitted using the 
described procedure.  The resulting curve of 
$-\log \mathcal{L}$ versus \plb, offset so the maximum value of
$\log \mathcal{L}$ is zero, is shown in Fig.~\ref{fig:result}(b)
indicating a measurement of
$$
\plb = -0.56^{+0.20}_{-0.13},
$$
and a value of zero polarization ruled out at the 95\% confidence
level (CL), considering only the statistical errors.
The $\chi^2$ per degree of freedom is 0.91.  

\newpage

\vfill
\begin{center}
\end{center}

\newpage

\section{Systematic uncertainties and consistency checks}

\indent
The sources of systematic error that have been considered
are summarized in Table~\ref{tab:syst}.

 \begin{table}[!htb]
  \begin{center}
    \begin{tabular}{|lr|c|}
      \hline
        \multicolumn{2}{| l |}{Source of Uncertainty} & 
         $\Delta \plb $ \\[2pt]
      \hline 
$E_{\nu}$ resolution     & & $\pm 0.02$ \\
$E_{\nu}$ reconstruction & & $\pm 0.05$ \\
$E_{\ell}$ scale and shape & & $\pm 0.03$  \\
Selection criteria            & & $\pm 0.02$ \\
Background fraction and shape & & $\pm 0.04$ \\
b fragmentation & & $\pm 0.03$ \\
$\Lambda_{\mathrm{c}}$ polarization & & $\pm 0.02$ \\
$\mathrm{b} \to \tau$ & & $\pm 0.01$ \\
Fitting method & & $\pm 0.03$ \\
Theoretical uncertainty (form factor  & & $\pm 0.03$ \\
modelling, QCD corrections, $m_{\mathrm{c}}/m_{\mathrm{b}}$) &  & \\ \hline
{\bf Total} & & $\pm 0.09$ \\ \hline
    \end{tabular}
  \end{center}
  \caption{\label{tab:syst} \sl
   Summary of systematic uncertainties in the measurement
of \plb.
  \normalsize }
\end{table}

The overall rate or efficiency for tagging $\Lambda \ell \pi^+$ 
combinations is not used in the fit but
any discrepancy in rate between data
and Monte Carlo prediction or distortions as a function
of $E_{\nu}$, $E_{\ell}$, or their ratio can potentially affect the result.
Control samples were used in the following 
to estimate these effects.

To test the Monte Carlo modelling of reconstructed neutrino energy resolution,
events with missing energy in the b-tagged Monte Carlo control 
samples with identified leptons
in the range corresponding to the missing energy in 
the $\lb \rightarrow \lc \ell \nu_{\ell}$
signal were studied.  Comparing the resolution $\sigma(E_{\nu})$
between the signal sample and the control sample 
in bins of $E_{\nu}$, differences of up
to 12\% were observed. Propagating this resolution uncertainty
into the PDF's for $E_{\nu} / E_{\ell}$ resulted in differences
of $\pm 0.02$ in \plb.

The linear correction to $E^{\mathrm{hemi}}_{\mathrm{miss}}$
used to reconstruct $E_{\nu}$ was varied within its uncertainties
in slope and intercept resulting in an observed variation
of $\pm 0.03$ in \plb.
Events with a lepton opposite a hemisphere failing a b-tag
were used as
an independent data control sample.
In this sample, the data and Monte Carlo distributions were
normalized to the same number of events and the ratio of the
resulting $E^{\mathrm{hemi}}_{\mathrm{miss}}$ distributions in 
bins of  $E^{\mathrm{hemi}}_{\mathrm{miss}}$
was fitted to a first-order polynomial.
The fitted slope was consistent with zero, but $\Lambda \ell \pi^+$
combinations were reweighted according to the one standard deviation
bounds on a possible slope.  
It was checked that a fit to a 
second-order polynomial resulted in no significant 
parabolic coefficient.  Similar results were observed with a
sample consisting of events 
with a lepton opposite a hemisphere passing the b-tag.  
Variations ranged up to $\pm 210$ MeV in the average value 
of $E_{\nu}$ and resulted in a change of $\pm 0.05$ in 
\plb.  This is assigned as the systematic error due to 
uncertainties in 
the $E_{\nu}$ reconstruction and its distribution.

The absolute scale of the momentum measurement of electrons and
muons was determined by comparing $\mathrm{e^+e^-}$ and 
$\mu^+\mu^-$ pair and photon conversion data with corresponding 
Monte Carlo simulation.
Lepton momenta in the data were rescaled by the observed 
uncertainty of $\pm 0.4\%$ resulting in a change in \plb\ of
$\pm 0.01$.
Possible mismodelling of the electron identification efficiency
as a function of $E_{\ell}$ was investigated by comparing the
relative rate in
multihadronic events of cleanly identified photon conversions with
Monte Carlo predictions as a function of electron energy.
Similar studies~\cite{muonid} were done for muons as a function
of muon momentum.  In both cases, relative variations did not
exceed 3\%, and Monte Carlo $E_{\ell}$ spectra 
were reweighted according to
the observed differences.  Refitting resulted in observed
changes of $\pm 0.03$ in \plb.
The uncertainty in the momentum resolution results
in a negligible effect on the polarization and the total assigned
systematic error due to uncertainties in the $E_{\ell}$ distributions
was $\pm 0.03$.

The selection criteria were varied within typical ranges of their
resolution as determined by signal Monte Carlo samples and the fit
repeated.  The observed variations of $\pm 0.02$ in \plb\ were
assigned as a systematic error.

The effect of uncertainties in the determination of the shape 
of 
the wrong-sign background PDF and the level of the 
background was assessed by varying each
component by its uncertainty indicated in Table~\ref{tab:back}.
Since the background due to a fragmentation $\Lambda$ baryon combined
with a lepton from b-baryon decay was predicted to
be $(21 \pm 3)\%$ of the wrong-sign sample,
the fit was also repeated after scaling down the background in 
the right-sign sample by this factor.
Lastly, instead of using the wrong-sign background PDF fit to the
number of wrong-sign combinations, the distribution of $E_{\nu}/E_{\ell}$
in the data wrong-sign combinations was simply subtracted from
the distribution of the right-sign combinations and the fit
to the signal PDF's repeated.  The observed changes from the
central value of \plb\ in these
tests were added in quadrature resulting in a systematic error
of $\pm 0.04$ assigned due to uncertainties in the background 
estimate.

The fragmentation parameter
$\langle x_E \rangle_{\mathrm{b}}$, defined as the average
fraction of the beam energy carried by a weakly decaying
b hadron, has been estimated to be
the same for baryons and mesons at the 1\% level~\cite{ratios1}.  
Varying the average \lb\ energy by the
measured errors on the average b-hadron energy~\cite{LEPcomb}
added in quadrature with this 1\% uncertainty results in a variation
of \plb\ of $\pm 0.03$ due to the residual logarithmic dependence
on fragmentation of the $E_{\nu}/E_{\ell}$ distribution~\cite{ratios2}. 

Polarization of the \lcp\ baryon can affect the acceptance for combinations in
the signal chain $\lb \to \lcp \ell^- \bar{\nu}_{\ell}$, 
$\lcp \to \Lambda \mathrm{X}$, $\Lambda \to \mathrm{p} \pi^-$
and also the lepton spectra from background semileptonic decays of 
\lc\ either produced in the above chain or via direct production
in $\mathrm{Z^0} \to c\bar{c}$ events.
In the first two cases, measured \lc\ decay parameters~\cite{PDG}
were used and $\Lambda \ell \pi^+$ Monte Carlo combinations 
involving \lc\ baryons entering the sample for the background  PDF
were reweighted (also see Ref.~\cite{lambcpol}) 
to correspond to variations of the
\lc\ polarization between 0 and $-1$.
In the latter case, the \lc\ polarization was varied between
0 and $-0.68$, the value for c quarks produced in  
$\mathrm{Z^0} \to c\bar{c}$.  The fitted value of \plb\ changed
by $\pm 0.02$, which is assigned as a systematic error due to this
effect.

The measured branching ratio for 
$\mathrm{Br(b} \rightarrow \tau 
\overline{\nu}_{\tau} \mathrm{X})$ was varied
within its uncertainties~\cite{btotau} and the level of transfer of
polarization of the tau chosen from an alternative
model~\cite{falktau} to reweight Monte Carlo background combinations.
The effect was small, and a systematic error of $\pm 0.01$ due to this
background source uncertainty was assigned.

Monte Carlo tests for possible biases in the fitting method have already
been 
described earlier. 
In addition, the binning of the spectra and the PDF's 
was varied from 30 bins to 20 and 40 bins. The range
of the fit was also changed from $ -2 \leq E_{\nu}/E_{\ell} \leq 4$
to $ -1 \leq E_{\nu}/E_{\ell} \leq 3$ and
the fit repeated.
Adding the observed variations in \plb\ in quadrature, a systematic
error of $\pm 0.03$ was assigned due to possible biases
introduced by the fitting method.

Many sources of uncertainty in the lepton and neutrino
energy spectra partially cancel in the ratio $E_{\nu} / E_{\ell}$.
For the input to the theoretical calculation of
these spectra for polarized b baryons
including QCD corrections~\cite{qcd2}, the
ratio $m_{\mathrm{c}}/m_{\mathrm{b}}$ was varied in the
range 0.22--0.39 and 
the strong coupling constant
$\alpha_S(m_{\mathrm{b}})$ in
the range 0.20--0.24~\cite{PDG}.
An alternative simple quark model~\cite{simple} using different
form factors in the polarized \lbz\ decay was also substituted. 
The signal Monte Carlo spectra
were weighted accordingly.
Different masses of b baryons and c baryon decay products
(e.g., \xb\ and \xc) were also used in the calculation of the
spectra. In all cases the fit was repeated and from the observed
variation in \plb, a systematic error of $\pm 0.03$ was assigned
due to theoretical uncertainties in the predicted
$E_{\nu} / E_{\ell}$ spectra.
   
The systematic uncertainties from each source were added in
quadrature to obtain an estimated total systematic error of
$\pm 0.09$ on \plb.
 
\noindent
{\bf Consistency Checks:} 
Further checks were performed to search for other 
systematic effects.
The cut on the minimum momentum of the electron or muon
was varied between 3.0 and 5.0~GeV, and the requirement
on the minimum $p_t$ 
changed between 0.6 and 1.0~GeV.  These
variations change the b quark content of the sample; in particular
the \lc\ and charm meson content increases for the smaller
$p_t$ cut.  The minimum momentum of
the $\Lambda$ candidate was varied in the range 3.0 to 5.0~GeV,
increasing and decreasing the background level of fragmentation $\Lambda$
baryons being combined with leptons.  In each case the analysis 
was repeated and \plb\ redetermined.  Observed variations were
consistent with the statistical errors on the uncorrelated fractions
of the different samples.

The fit was performed separately for data collected in
1990--1993 and for 1994--1995, and also separately for
$\Lambda \mathrm{e}^- \pi^+$ and $\Lambda \mu^- \pi^+$ combinations.
All results were statistically consistent with each other and with the
central value from the full data set.
The observed rate of $\Lambda \ell^- \pi^+$ combinations were
also consistent in these different samples, and the
derived value
$f(\mathrm{b} \to \lb ) \cdot 
{\mathrm{Br}}(\lb \to \Lambda \ell \bar{\nu}_{\ell} \mathrm{X}) =
(2.78 \pm 0.16) \times 10^{-3}$ (statistical error only) 
agrees with previous OPAL measurements~\cite{lbcorrel,semibr} within 
errors.

A fit was also made to the reconstructed
$E_{\nu}/E_{\ell}$ distribution in the inclusive lepton control sample from 
the data.  This large statistics sample consisted of events
with an identified lepton in the hemisphere opposite
to a b-tagged hemisphere.  No wrong-sign background subtraction
was performed, and 
a longitudinal polarization of 
$\pl = -0.043 \pm 0.019 \thinspace \mathrm{(stat.)}$ was
measured. 
Although B mesons are expected to be unpolarized
(the ground states are unpolarized and for spin-1 states
the polarization is undetectable), a fraction  
$(10.1^{+3.9}_{-3.1})\%$~\cite{PDG} of the b hadrons are predicted 
to be b baryons in this sample, and there will also be a 
contamination of polarized c baryons that decay weakly.
In addition, the smaller expected semileptonic branching ratio
for b baryons compared to the other b hadrons~\cite{semibr,PDG} 
would result
in further reduction of total observed polarization.
Although the total observed polarization is not necessarily 
expected to be zero,
it should be small, as observed.

A further cross check was made with a data sample
containing candidates for the exclusive decay
$\mathrm{B}^0 \to \mathrm{D}^{*-} \ell^+ \nu_{\ell}$ selected
from events where a $\mathrm{D}^{*-}$ and a lepton of opposite
charge were found in the same jet~\cite{barberio}.  
A value consistent with zero,  $\pl = 0.07^{+0.13}_{-0.12}$,
was found as expected for a sample with a b hadron content
almost exclusively from B~mesons and much less contamination
from possibly polarized heavy baryons.
 

\section{Discussion and summary}

\indent
The average polarization of b baryons in $\mathrm{Z^0}$ decays at
OPAL has been measured to be:
$$
\plb  = -0.56^{+0.20}_{-0.13} \pm 0.09,
$$
where the first error is statistical and the second systematic. 
This level of polarization is larger than, but consistent with, the
published measurement by the ALEPH Collaboration~\cite{ALEPHpol} of 
$\plb  = -0.23^{+0.24}_{-0.20} \thinspace (\mathrm{stat.})
            \thinspace ^{+0.08}_{-0.07} \thinspace (\mathrm{syst.})$.


Including systematic errors, this OPAL measurement implies
bounds on the longitudinal polarization of b baryons of
$-0.13 \geq \plb \geq -0.87$ at 95\% CL, therefore disfavoring
full observed average polarization of $-0.94$.  This is the first
measurement to exclude zero polarization of b baryons at
larger than 95\% CL providing direct evidence that the
b quark is longitudinally polarized in the decay
$\mathrm{Z^0} \to \mathrm{b\bar{b}}$.

A simple model~\cite{bmeson} can be used to predict the total
observed b baryon polarization after depolarization of those
b baryons proceeding through intermediate states involving strong decays:
$\mathrm{b} \to \Sigma_{\mathrm{b}}^{(*)} \to \lb$.
The prediction depends on a parameter $A$ that is the relative
probability of producing a spin $S = 1$ diquark as opposed to
a $S = 0$ diquark, and a parameter $\omega_1$ that is the probability
that the spin 1 diquark has angular momentum component 
$j_3 = \pm 1$ along the fragmentation axis.  For $\omega_1 = 0$
(alignment suppression) and the default JETSET 
parameter\footnote{In the JETSET package, 
$A \approx 9 \cdot \mathrm{PAR(4)}$.}  for
$A$, a value of $\plb = -0.68$ is predicted.
Varying $\omega_1$ to 0.66 (isotropic diquark spin distribution) and
$A$ within bounds suggested by measurements~\cite{strange} gives a
range of predictions between $-0.54$ and $-0.88$.
In this model, the measured value of \plb\ is consistent with
no depolarization during fragmentation.  It is also consistent
with the inclusive $\Lambda$ measurement of OPAL~\cite{inclla}
that also found no evidence of polarization loss in hadronization.

\section*{Acknowledgements}
\noindent

We thank M.\ Je\.{z}abek for providing
the {\textsc{Fortran}} program of the calculations of 
the neutrino and lepton spectra as described in Ref.~\cite{qcd2}.
We particularly wish to thank the SL Division for the efficient operation
of the LEP accelerator at all energies
 and for their continuing close cooperation with
our experimental group.  We thank our colleagues from CEA, DAPNIA/SPP,
CE-Saclay for their efforts over the years on the time-of-flight and trigger
systems which we continue to use.  In addition to the support staff at our own
institutions we are pleased to acknowledge the  \\
Department of Energy, USA, \\
National Science Foundation, USA, \\
Particle Physics and Astronomy Research Council, UK, \\
Natural Sciences and Engineering Research Council, Canada, \\
Israel Science Foundation, administered by the Israel
Academy of Science and Humanities, \\
Minerva Gesellschaft, \\
Benoziyo Center for High Energy Physics,\\
Japanese Ministry of Education, Science and Culture (the
Monbusho) and a grant under the Monbusho International
Science Research Program,\\
German Israeli Bi-national Science Foundation (GIF), \\
Bundesministerium f\"ur Bildung, Wissenschaft,
Forschung und Technologie, Germany, \\
National Research Council of Canada, \\
Research Corporation, USA,\\
Hungarian Foundation for Scientific Research, OTKA T-016660,
T023793 and OTKA F-023259.\\


\begin{thebibliography}{99}
\bibitem{popcorn}    B.~Andersson, G.~Gustafson, and T.~Sj\"ostrand,
                     {\sl Physica Scripta} {\bf 32} (1985) 574. \newline
                     A popcorn model has been implemented in JETSET,
                     where the frequency for the process
                     baryon-meson-antibaryon is regulated in JETSET by
                     the parameter PARJ(5) (with default value 0.5),
                     such that the probability
                     is given by PARJ(5)/(0.5 + PARJ(5)).
\end{thebibliography}
\end{document}